\documentclass[aps,pra,superscriptaddress,numerical,showkeys,floatfix,longbibliography,preprint, footinbib]{revtex4-2}
\usepackage{epsfig,amsmath,amssymb,txfonts,hyperref,multirow, bbold} 
\usepackage[version=4]{mhchem}
\usepackage{xcolor}
\usepackage[normalem]{ulem}
\usepackage[figuresleft]{rotating}
\usepackage{bm}

\newcommand{\be}{\begin{equation}}
\newcommand{\ee}{\end{equation}}

\newcommand{\Fig}[1]{Fig.~\ref{fig:#1}}

\newcommand{\Eqn}[1]{Eqn.~\ref{eqn:#1}}

\newlength{\wholefigwidth}
\newlength{\halffigwidth}
\begin{document}

\title{Explainable Machine Learning for Oxygen Diffusion in Perovskites and Pyrochlores}
\author{Grace M. Lu}
\author{Dallas R. Trinkle}
\email{dtrinkle@illinois.edu}
\affiliation{Department of Materials Science and Engineering, University of Illinois at Urbana-Champaign, Urbana, Illinois 61801, USA}

\date{\today}
\begin{abstract}
Explainable machine learning can help to discover new physical relationships for material properties. To understand the material properties that govern the activation energy for oxygen diffusion in perovskites and pyrochlores, we build a database of experimental activation energies and apply a grouping algorithm to the material property features. These features are then used to fit seven different machine learning models. An ensemble consensus determines that the most important features for predicting the activation energy are the ionicity of the A-site bond and the partial pressure of oxygen for perovskites. For pyrochlores, the two most important features are the A-site $s$ valence electron count and the B-site electronegativity. The most important features are all constructed using the weighted averages of elemental metal properties, despite weighted averages of the constituent binary oxides being included in our feature set. This is surprising because the material properties of the constituent oxides are more similar to the experimentally measured properties of perovskites and pyrochlores than the features of the metals that are chosen.
The easy-to-measure features identified in this work enable rapid screening for new materials with fast oxide-ion diffusivity.
\end{abstract}

\keywords{machine-learning; oxides; diffusion}

\maketitle

\section{Introduction}
\label{sec:intro}
Fast oxide-ion conductors have been studied due to their applications in solid oxide fuel cells (SOFC), either as electrolytes if they are electronic insulators \cite{ Fergus2006, Wang2019, Orera2010, boivin1998recent, Goodenough2003, Ormerod2003, Jacobson2010, Gauckler2004}, or as cathodes if they are mixed electronic and oxide ion conductors \cite{Goodenough2003, Ormerod2003, Gauckler2004, Jacobson2010, Mahato2015, Steele2001}, oxygen sensors \cite{Steele1992}, oxygen permeation membranes \cite{Steele1996, Sadykov2019}, or as catalysts for syngas production \cite{Sadykov2019, Gellings2000}. For SOFC electrolytes, their ionic conductivity is the most important material property that determines the operating temperature and power output. However, significant oxide mobility (greater than 10$^{-2}$ S cm$^{-1}$) has only been observed at high temperatures between 800 and 1000$^{\circ}$C 
\cite{boivin1998recent}. These high operating temperatures lead to thermal and chemical instability, so lowering the operating temperature is necessary to improve SOFC lifetimes and efficiency \cite{Goodenough2003, Wachsman2011}. Two crystal structures of oxide-ion conductors that have been observed to have high oxide-ion mobility are perovskites and pyrochlores \cite{boivin2001structural}. To predict new promising SOFC electrolytes, it is vital to obtain a quantitative understanding of oxygen ion diffusion in these materials to enable rational design of new materials.

Earlier models for predicting perovskite conductivities have found a high importance for bond character and structure. While more direct surrogates for the bond character have been used with ab initio calculations, models fit to experimental data typically rely on elemental electronegativities. 
One of the first models for perovskite conductivities by Xu et al. \cite{xu2009two} fit a quadratic relationship between the log of the oxygen conductivity and the ratio of the O-O bond charge population with the O-O bond length for five DFT-calculated oxide ion conductivities in perovskites at 1073 K. 
They hypothesize that the ratio between the O-O charge population and the bond length is a measure of the bonding strength in the edge of the BO$_6$ octahedron, but accurate calculation of this ratio requires expensive ab initio calculations. 
To improve the generalizability, they further fit a support vector machine to logarithmic experimental oxide ion conductivities, using six features: tolerance factor, electronegativity difference between A or B and oxygen, charge A or B, and $\Delta EDB/R_{R+O} = \frac{\textrm{Electronegativity}_{\textrm{O}} - \textrm{Electronegativity}_{B}}{\textrm{Radius}_\textrm{O} + \textrm{Radius}_\textrm{B}}$. Their work was followed by Zhang and Xu \cite{ZHANG2022} who fit a Gaussian process model using the same features and dataset. While no formal feature importance analysis was done, they fit six models with half of their original features. There were four pairs of features with a larger than 0.6 correlation with another (electronegativity difference A and charge A, electronegativity difference B and charge B,  charge A and $\Delta EDB/R_{R+O}$, and charge B and $\Delta EDB/R_{R+O}$). For each of the six models, they chose to remove three features---one from each of the pairs---and found that the training RMSE increased by at least 0.32 after removing features compared to the original RMSE of 0.0295 when including all features. The worst performing model had an RMSE of 0.931 after removing charge A, charge B, and $\Delta EDB/R_{R+O}$. They suggest that the tolerance factor describes the structural stability, while the other five describe the electronic structure, particularly in the BO$_6$ octahedron.
While these models succeeded in having accurate predictions on training data, they rely on a small set of crafted features, and thus, may be missing previously unknown relationships between material features and the conductivity.
A more comprehensive paper by Priya and Aluru \cite{Priya2021} fit a XGBoost model to experimental perovskite conductivities with a much larger set of 111 features, and found that the two most important features were the B-site minimum electronegativity and the average B-site ionic radius. They suggest that larger electronegativities lead to higher metallic characteristics, leading to larger conductivities. 
Because only 11\%\ of Priya and Aluru's dataset consisted of primarily oxide ion conductors, it is unclear if their findings can be directly applied to oxide ion diffusion or if their conclusions are more applicable to mixed or electronic conduction.

Oxygen diffusion in pyrochlores is more poorly understood than in their perovskite counterparts, with no previous machine learning model for predicting diffusion in pyrochlores. Disagreement between DFT calculations and experimental measurements occurs for the diffusion mechanism. Pirzada et al. performed molecular dynamics simulations of 54 pyrochlores using an empirical force field \cite{PIRZADA2001201}, which was later followed up on by Li et al.'s DFT calculations \cite{LI2018255}. Both calculations concluded that the single vacancy had a lower formation energy than the split vacancy for the titanates. For Zr-containing pyrochlores, smaller A-site cation radii resulted in a more energetically favorable split vacancy mechanism. Experimental characterization of Yb$_2$Ti$_2$O$_7$, which both calculations predicted to have a single vacancy mechanism, used neutron and X-ray diffraction patterns to find that oxide diffusion occurs along both the [100] and [110] directions in the pyrochlore, which is only possible with the split vacancy mechanism \cite{uno2018experimental}. Ugawa et al. used ab initio molecular dynamics to propose a two-step cooperative mechanism for Yb$_2$Ti$_2$O$_7$ where two neighboring oxygen atoms migrate together and suggest that both single and split vacancies coexist at high temperatures \cite{matsumoto2021cooperative}.

Multiple machine learning models have previously been fit to vacancy formation energies in oxides. Wan et al. fit multiple machine learning models using six designed features and found that the two features with the highest correlation with the vacancy formation energy were the differences in atomic electronegativity and the proportion of oxygen electrons in the metal oxide \cite{wan2021data}. This conclusion was further supported by Deml et al. who constructed simple linear models to predict DFT-calculated vacancy formation energies in oxides and found that the most important features were the strength and nature of the chemical bonding \cite{deml2015intrinsic}. Similarly, Wexler et al. concluded that a simple linear model with only two features: crystal bond dissociation energy and the reduction potential could be used to accurately predict oxygen vacancy formation energies\cite{wexler2021factors}.
Of particular interest is Baldassari et al. fit random forest models to 2677 DFT-calculated vacancy formation energies in 1157 oxides, of which the most prevalent structures were A$_2$BB'O$_5$ layered perovskites, perovskites, and pyrochlores \cite{baldassarri2023oxygen}. They noted that including the minimum and weighted average of binary oxygen unrelaxed vacancy formation energies with the MAGPIE features halved the test mean absolute error, showing the benefits of including information about the constituent binary oxides, not just the elemental properties from MAGPIE.

Previous machine learning models for solute diffusion in metals \cite{Wu2017, Lu2019} and hydrogen diffusion in metals \cite{lu2023explainable}  have shown that the individual feature importances often disagree between different models and cross-validation techniques as a result of classic machine learning techniques being unable to accurately compensate for correlations between features. 
Our proposed grouping algorithm uses these large correlations between features to group them, allowing each group's importance to be combined and analyzed in unison, which leads to consistent feature importances between machine learning models. This method, however, has only previously been applied to diffusion in metals and random binary alloys \cite{lu2023explainable}, where the material's true physical properties can be more accurately represented using weighted averages of the elemental properties unlike the oxides.

In this work, to predict experimental oxygen diffusion activation energies in perovskites and pyrochlores, we fit seven different machine learning models: recursive feature machine, $k$-nearest neighbors, Gaussian process, random forest, gradient boosting tree, Bayesian Ridge, and linear models. We measure their performance using three different test-train splits (all training, leave-one-out, and 80--20\%\ cross-validation) to understand the relationships between the material properties and the activation energy. Using these models and our feature grouping method, we analyze their feature importances and obtain physical insight into the diffusion of oxygen in these oxides. These findings point to the importance of elemental electronegativities in understanding and designing new oxides with the ideal oxygen diffusivities.

\section{Methods}
\label{sec:methods}

\subsection{Database}

\begin{figure*}[htbp]
 \centering\includegraphics[width=\wholefigwidth]{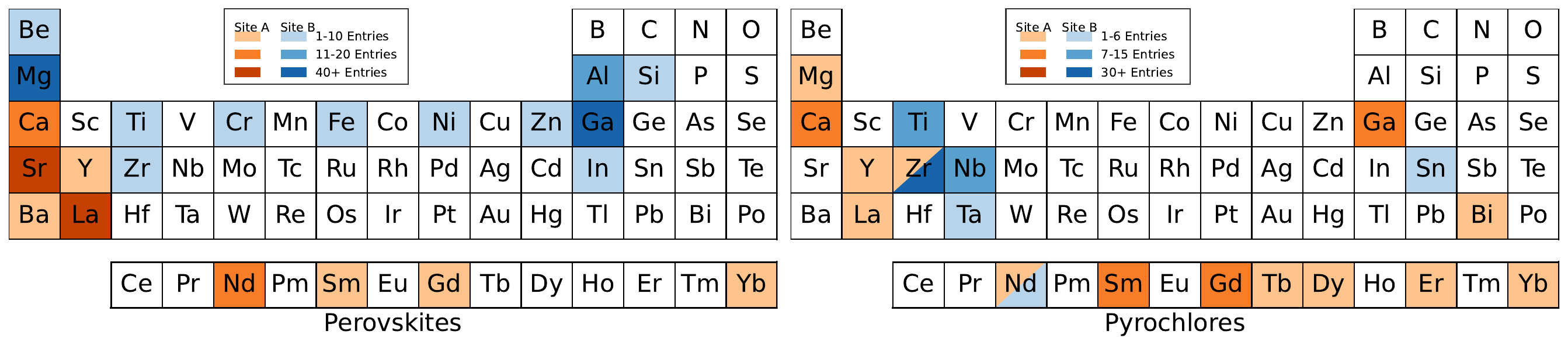}
 \caption{Count of the elements included in the perovskites (left) and pyrochlores (right) database split by site. For perovskites, La is the most common element contained in 62 perovskites, followed by Ga with 60 and Sr with 57. Zr is the most common element in the pyrochlore database, occurring in the B site for 34 pyrochlores.}
 \label{fig:perovdatabase}
\end{figure*}

We construct an experimental database of oxygen ionic conductivity activation energies for 76 perovskites \cite{ishihara1994doped, iwahara1992oxide, ishihara1994oxide, lybye2000conductivity, huang1998superior, kharton2000oxygen} and 42 pyrochlores \cite{Anithakumari201697566, SIBI20091164, HAGIWARA2014551, gill2012ionic, XIONG20118392, LIU2011385, DIAZGUILLEN20082160, Radhakrishnan2912, ALMEIDA20121275, KRASNOV2017118, Burgraaf1981, Tuller1994, Kramer1994, vanDijk1983913, Moon_Tuller_1988} in \Fig{perovdatabase}. Most experimental measurements of the oxygen ionic conductivity were performed through electrical conductivity measurements. For Gd$_2$Zr$_2$O$_7$, the most frequently measured pyrochlore, the experimental measurements varied from 0.73--0.90 eV \cite{vanDijk1983913, Moon_Tuller_1988, Burgraaf1981}. These experimental measurements were heavily focused on specific elements with 62 perovskites containing La and 34 pyrochlores containing Zr. There is less elemental variety on the B-site for pyrochlores, with only 6 elements appearing in total, while 14 different elements can appear on the A-site. We find that the elements that appear on either site overlap between perovskites and pyrochlores, as transition metals tend to appear on the B-site and lanthanides appear on the A-site. There are 4 perovskites with only 2 unique metal elements (the A and B site have only one type of atom), and 61 have 4 elements. For the perovskites, the maximum number of elemental types on a site is 2. In the pyrochlore database, there are 12 entries with only 2 unique elements, but there are 4 entries with 5 unique elements where 2 appear on the A-site and 3 appear on the B-site. Thus, the vast majority of our entries include some form of elemental disorder on the cation sublattice. For the perovskites, on average, only 10\% of the dopant element is present on either the A or B site. Only one perovskite (Ti$_{0.5}$Al$_{0.5}$CaO$_3$) has 50\% occupation of both Ti and Al on the A site. For pyrochlores, this is true for 11 different materials. For both of these oxide structures, the B-site is closer than the A-site for oxygen sites involved in the diffusion mechanism. In perovskites, the B-site is the nearest neighbor for the single oxygen site. In pyrochlores, with their more complicated crystal structure, the occupied 48f oxygen site is nearest neighbors with 2 A-sites and 2 B-sites, but the vacant 8b site is nearest neighbors to 4 B-site atoms. The 8b site is theorized to be involved in oxygen diffusion in pyrochlores, as mobile vacancies are created on the 48f site by exciting oxygen into the 8b site \cite{wilde1998defects}.

\subsection{Features}

\begin{figure*}[htbp]
 \centering\includegraphics[width=\wholefigwidth]{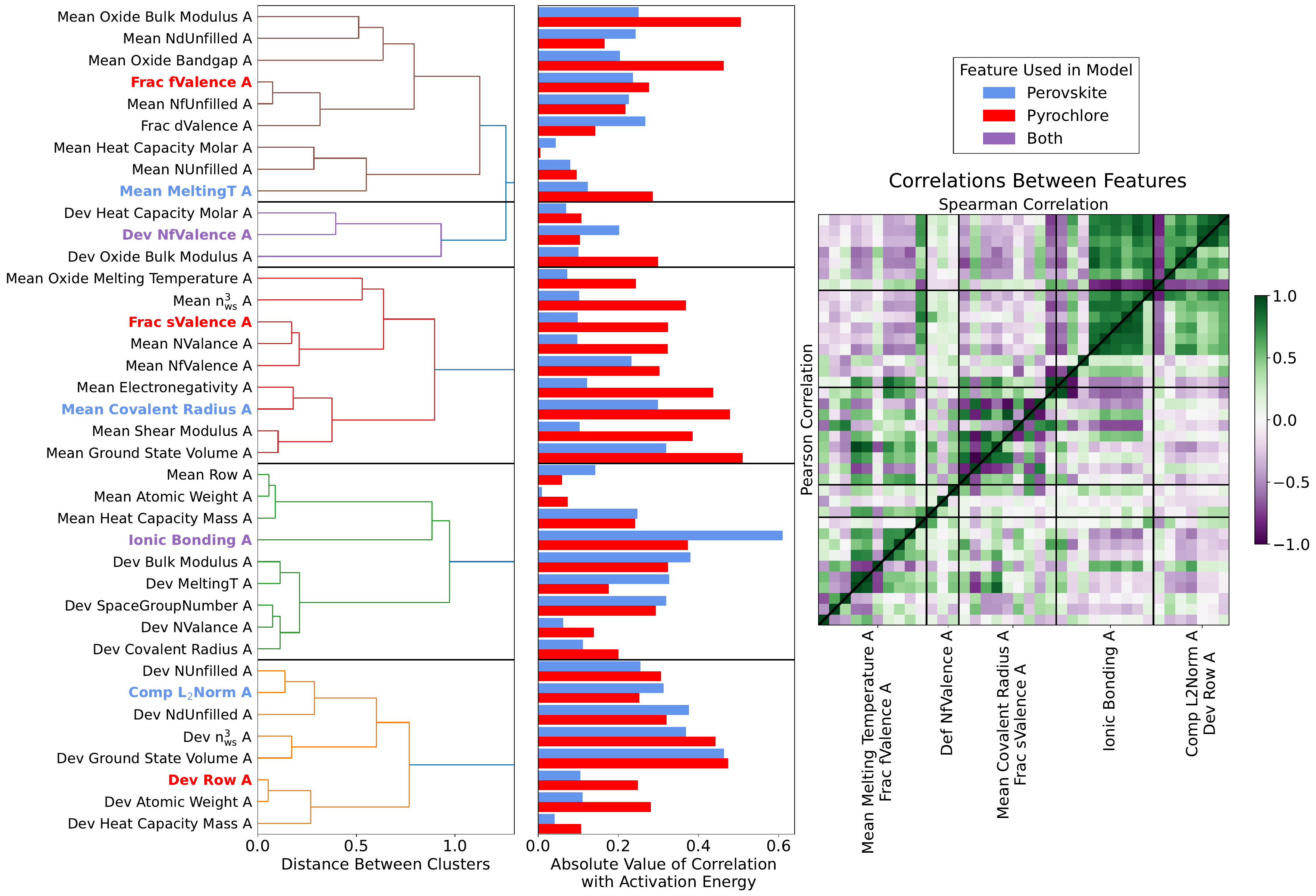}
 \caption{Hierarchical grouping of site A features using the Ward variance minimization. For all oxides, we use the weighted average (mean) of the elemental features by atomic composition without oxygen and the standard deviations (dev) of the MAGPIE features. The six features with the oxide label are combinations of the oxide building blocks. We use the Spearman rank correlations to define the distance between features, shown in the upper triangle of the right panel; the lower triangle provides the Pearson correlations. Using a greedy algorithm that groups the two closest features at every step, Ward variance minimization create groups of correlated features in the left panel. We chose a distance threshold of 1.2 to define groups, which are separated by black lines. The representative features for each group are bolded in the labels on the left and used as labels in the correlation plot. These are chosen through a greedy algorithm that chooses each feature with the highest correlation with a residual of a linear model with the activation energy and all previously chosen features.}
 \label{fig:corrA}
\end{figure*}

\begin{figure*}[htbp]
 \centering\includegraphics[width=\wholefigwidth]{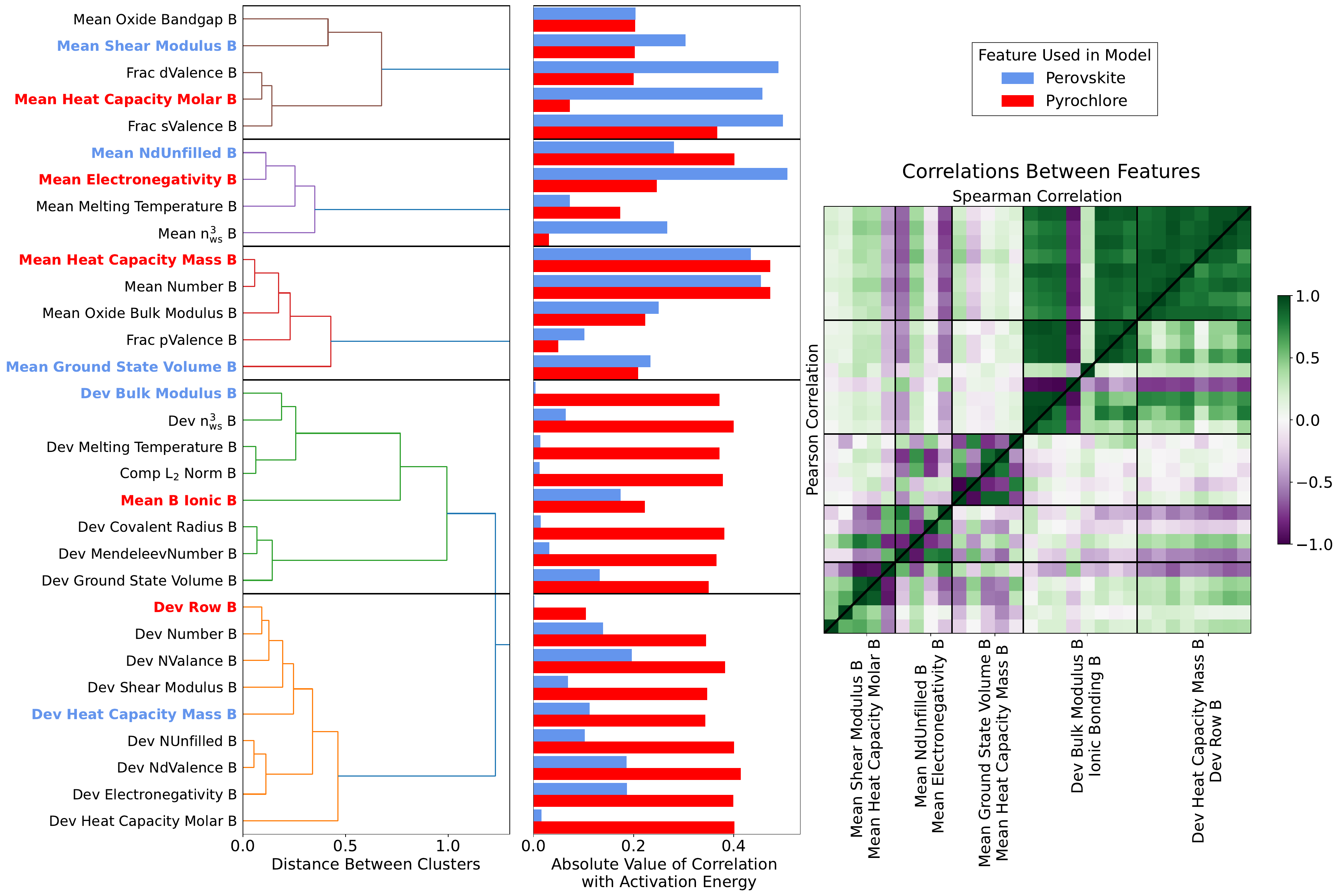}
 \caption{Hierarchical grouping of site B features using the Ward variance minimization. For all oxides, we use the weighted average (mean) of the elemental features by atomic composition without oxygen and the standard deviations (dev) of the MAGPIE features. The six features with the oxide label are combinations of the constituent binary oxides. We use the Spearman rank correlations to define the distance between features, shown in the upper triangle of the right panel; the lower triangle provides the Pearson correlations. Using a greedy algorithm that groups the two closest features at every step, Ward variance minimization creates groups of correlated features in the left panel. We choose a distance threshold of 1.2 to define groups, which are separated by black lines. The representative features for each group are bolded in the labels on the left and used as labels in the correlation plot. These are chosen through a greedy algorithm that chooses each feature with the highest correlation with a residual of a linear model with the activation energy and all previously chosen features. Site B features have a more pronounced block structure than their site A counterparts, and there are much weaker correlations between the top two and bottom two feature groups.}
 \label{fig:corrB}
\end{figure*}

Our material features are shown in the left column of \Fig{corrA} and \Fig{corrB}; they consist mainly of elemental weighted averages and standard deviations split by site, but also include three ``oxide properties", one experimental condition (the partial pressure of oxygen), and the space group number of the oxide. Most of our features were generated using MAGPIE, which calculates elemental weighted averages and standard deviations of the elemental properties \cite{Ward2016}. To improve the specificity of our feature importance analysis, we create a separate feature for each site, which we have labeled by including either A/B at the end of the feature name. Note that neither the elemental mean nor standard deviation (labeled as Dev in the feature names) are accurate representations of the perovsksite or pyrochlore's properties, but act as easier-to-calculate surrogates. While the ionic radii is a feature that is used frequently to predict features in oxides, we choose not to include it in favor of including the periodic table positions, atomic radii, and electronegativities. This is because in our database, we have mixed charge states due to the inclusion of aliovalent doping. In these cases the ionic radii can change significantly. Likewise, the Goldschmidt tolerance factor, which is a function of the ionic radii, cannot serve as a reliable indicator of the material properties. Additionally, while a similar tolerance factor has been formulated for pyrochlores \cite{song2020tolerance, saha2022empirical}, they are hampered by the pyrochlore structure's ability to accommodate defects, and thus, none are predictive for stability \cite{fuentes2018critical}. In constructing our all but one of the features, we ignore oxygen's properties in the calculation, as it is included in a fixed stoichiometric ratio. Thus, including oxygen is equivalent to adding a constant to all of our values. The one feature where oxygen is included in the calculation is the Ionic Bonding A(B) feature, which is defined as \begin{equation}\sum_{i\in A}\sum_{j} x_i x_j (1 - \exp({-0.25( \chi_i-\chi_j)^2)}
 \label{eqn:ionicbonding}\end{equation}
where $x_i$ is the atomic fraction and $\chi_i$ is the electronegativity, and measures the strength of bonding between the A site cations and all other elements. Pauling's ionic bonding feature is an empirical measure based on the relationship between the electronic dipole moment of diatomic molecules and the electronegativity differences\cite{pauling1960nature}. It is also the only feature that explicitly combines A and B site features.
The ``oxide properties" (bulk modulus \cite{deJong2016}, bandgap \cite{matproject}, and melting temperature \cite{Haynes}) were constructed by splitting each oxide into its constituent binary oxides, then taking the mean and standard deviation of the binary oxides' properties. If a metal commonly forms multiple oxides, the oxidation state was chosen to match with the perovskite or pyrochlore. These ``oxide features'' should serve as  better estimations of the real properties of the perovskites or pyrochlores.  
The partial pressure of oxygen ($p_{\text{O}_2}$ ) denotes the atmospheric conditions where the experimental measures were performed. For values in our database, the $p_{\text{O}_2}$  is split into two categories: standard atmosphere conditions (20\%) and low oxygen (\textless 0.01\%). While 79\% of perovskites were measured in low oxygen atmospheric conditions, only 2 pyrochlores were not measured in standard atmospheric conditions. 
We perform a one-hot encoding for the space group number. While all pyrochlores have the same space group number (227), there are four different space groups for perovskites (62, 74, 167, 221). 

To reduce the total number of features, we build groups using the Spearman correlations between features and chose one feature from each of the groups as shown in \Fig{corrA} and \Fig{corrB}. Because there are non-negligible correlations between the A and B site features, we find that grouping all features simultaneously led to mixed feature groups that contain both A and B site features. To disambiguate elemental effects, we choose to split the A and B site features before grouping our features. Because the $p_{\text{O}_2}$  and one-hot-encodings of the space groups are not A or B site features, we do not include them in the grouping process. These groups are constructed with the entire perovskite and pyrochlore database, and the groups are the same for all machine learning models. We note that the B site features have larger correlations with other B site features than A site features. This is likely due to more elemental variety on the A site, particularly for pyrochlores where only 5 elements are on the B site versus 14 on the A site.
We use a cutoff of 1.2 to define the groups, which is chosen so that the number of groups was identical for the A and B site.
This cutoff leads to 5 groups for each site, for a total of 10. 
The machine learning models are given a single representative feature from each group, along with the $p_{\text{O}_2}$  and space group numbers. We train three separate groups of models on different subsets of our database: perovskites and pyrochlores, perovskites only, and pyrochlores only. Each model has a differing number of total features, which depends on the total number of space groups present in that subset of the database: 15 (perovskites and pyrochlores), 14 (perovskites only), and 10 (pyrochlores only). The $p_{\text{O}_2}$  and space group numbers are each treated as a group with only one feature. 
In order to choose the representative feature, we first choose the feature with the largest absolute value of the Pearson correlation with the activation energy. Then, using a greedy algorithm, we select the next features to have the largest correlation with the residual of a linear model fit with all previously-chosen features.  The representative feature vary between the three different models groups. For the linear model, the number of features are truncated to minimize the 80--20\% testing RMSE. The features are added in the same order that our representative features were chosen. The linear model receives a different number of features for each of the three database subsets: all (9), perovskite only (5), and pyrochlore only (2). 
We find that, generally, the space group numbers are late in the process. For the combined perovskite and pyrochlore model, no space group number is chosen in the first ten steps. For the perovskite only model, space group 167, which separates hexagonal perovskites from cubic ones, is the sixth feature selected by the greedy algorithm. The low selection rate is due to higher correlations between the space groups and the other representative features, which arises because these space group numbers were not included in the original grouping algorithm. These space group features are the only features with a larger than 0.6 Pearson correlation with any other representative feature.

\section{Results}
\subsection{Model Fitting}
\label{sec:model}

In this work, we fit seven models, which are suitable for predictions on smaller datasets: recursive feature machines (RFM) \cite{radhakrishnan2022mechanism}, $k$-nearest neighbors (KNN) \cite{cover, fix}, Gaussian process (GPR) \cite{rasmussen1997evaluation, williams1995gaussian},
random forest (RF) \cite{breiman2001random}
, gradient boosting (GBR) \cite{friedman2001greedy}, Bayesian Ridge (BR) \cite{bishop2006pattern, mackay1992bayesian, tipping2001sparse}, and a linear model. 
The RFM, GPR, and BR models are kernel-based, the RF and GBR consist of ensembles of decision trees, and both the BR and linear models fit linear functions of the features to the data. The RFM is a kernel machine that does not have a fixed kernel. Instead, the RFM recursively learns a feature matrix $M$ that is incorporated into the kernel $K$. This kernel is an extension to the Laplace kernel and is defined as $K(x,z) =\exp{(-\gamma (x-z)^TM(x-z))}$ where $\gamma > 0$ is a learnable constant, and $x, z$ are data points. The RFM has an in-built feature selection method, so we include all features without downselection. The RFM converged the feature matrix with 5 steps of iteration for a RMSE difference of 3 meV. 
The two Bayesian models, GPR and BR, give an error estimate that arises from their inclusion of uncertainty, and the BR is a simplification of the GPR with a less-flexible kernel. For the GPR, we chose the kernel form to be a sum of two kernels: the White and RBF where the White kernel models random noise. Inclusion of the White kernel allows us to explicitly model any noise in the experimental predictions. The KNN model does linear interpolation between the closest $n$ neighbors, and is the only model that does not explicitly fit a function.
A grid search optimizes the hyperparameters for our models using the cross validation testing error for a 80\% random split as an objective function. We optimized these hyperparameters for the following three models: KNN (number of neighbors, weighing scheme), RF (number of estimators, maximum depth), and GPR (learning rate, number of estimators, maximum depth). For the other two models (RFM, Bayesian Ridge), there are no hyperparameters to optimize. We find that the choice of hyperparameter has a small effect on the RMSE, as when we replaced the optimal hyperparameter with those optimized for our previous hydrogen activation energy dataset \cite{lu2023explainable} for the $k$-nearest neighbors, random forest, and gradient boosting tree models, the RMSEs only increase by less than 5 meV, despite the different features and training data used when choosing those sub-optimal hyperparameters. The other four models do not have hyperparameters to fit. To judge the performance of all seven models, we use three different test-train splits of our data: all training, leave-one-out, and an 80--20\% random split, and for the 80--20\% random split, we report the final RMSE over 200 different random splits.

\begin{figure*}[htbp]
 \centering\includegraphics[width=\wholefigwidth]{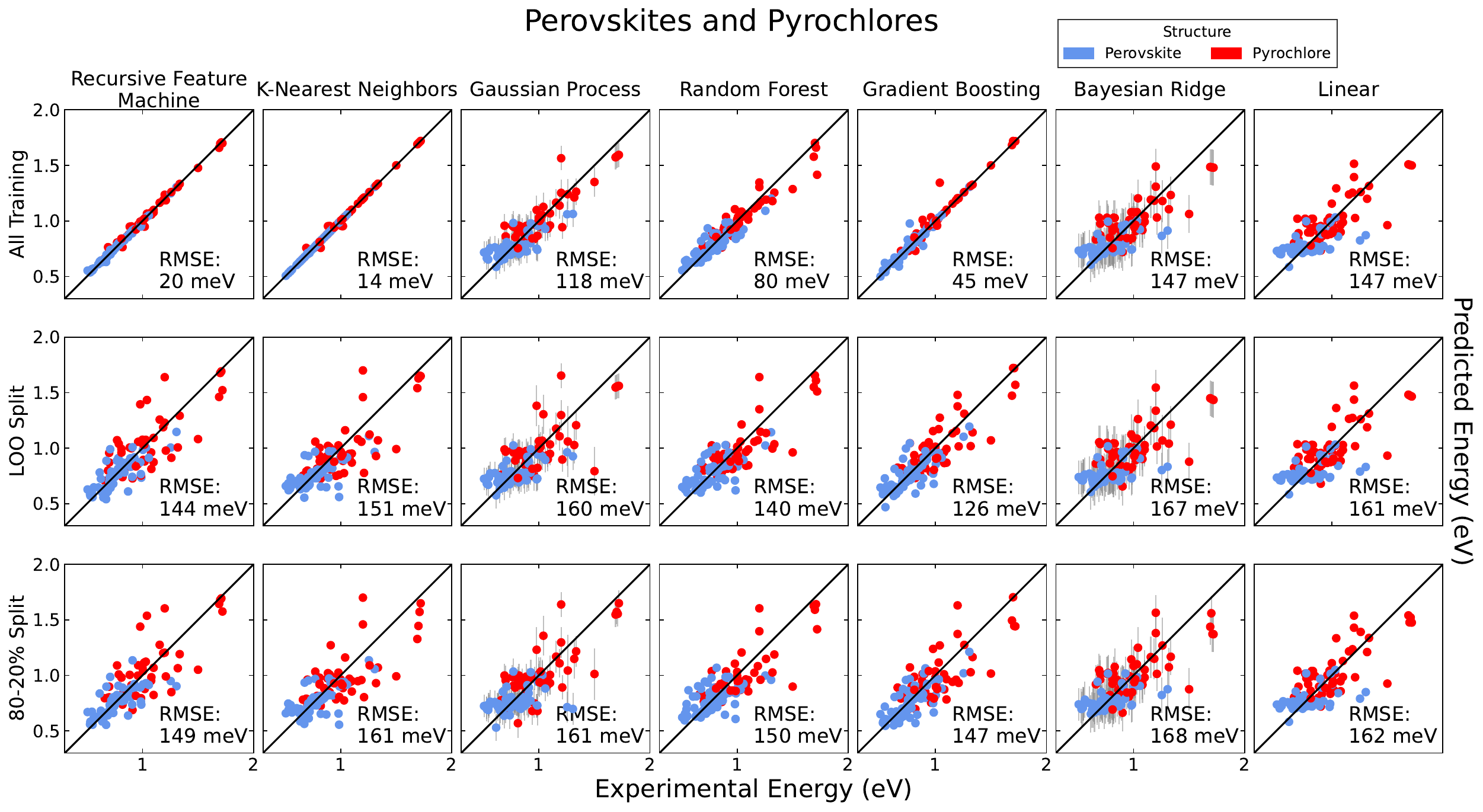}
 \caption{Predicted vs. experimental activation energies for oxygen in both perovskites and pyrochlores. In the middle and bottom rows, we plot only predictions on the test set over all cross-validation splits, and no training data is shown. The color denotes the crystal structure of the oxide. The two linear models, Bayesian Ridge and linear, have comparable RMSEs (162--168 meV) compared to the other five non-linear models (147--161 meV). This is despite major differences in the features and diffusion mechanism between the two crystal structures, so we can conclude that the differences can be accurately represented by a linear relationship.}
 \label{fig:all_models}
\end{figure*}

\Fig{all_models} shows the predictions on the combined perovskite and pyrochlore databases; we find that the two linear models, Bayesian Ridge and linear, have only slightly larger RMSEs (162--168 meV) compared to the others (147--161 meV) showing a fundamental similarity in the feature-property relationships for perovskites and pyrochlores that can be represented by a linear function. 
This is despite significant differences between the two oxide types. If we perform a two-sided Student's $t$-test comparing the perovskite and pyrochlore data to determine whether there is a statistically significant difference between the two groups, we find that almost all features, except for the Dev Bulk Modulus B, have a $p$-value that is larger than 0.05, which suggests a significant difference between the pyrochlore and perovskite features. This is to be expected as there is a large difference in the elements that form perovskites or pyrochlores, c.f. \Fig{perovdatabase}. 
The feature that explicitly splits the database into perovskites and pyrochlores (space group 227), however, is the second-to-last chosen feature, only ahead of space group 167, by the feature selection greedy algorithm, and thus, has one of the lowest correlations with any of the residuals. 
This is due to its high correlation with other features: it has a larger than 0.7 Pearson correlation with both the $p_{\text{O}_2}$  and Mean $n_{Ws}^3$ B, which were included as the first and fifth features using the representative feature greedy algorithm. The strong correlation between space group 227 and $p_{\text{O}_2}$  is due to differing experimental conditions between perovskites and pyrochlores; while 63 out of 76 perovskites were measured at $p_{\text{O}_2}$  less than 20\%, only 2 out of 42 pyrochlores were. Similarly, pyrochlores have a much larger value of the Mean Oxide Melting Temperature B compared to the perovskites.
Thus, even though the feature that directly measures the structure (space group 227) is not important for our models, the models are able to differentiate between them due to the other features. While the models can take into account a more complicated relationship than just a constant due to the differences in the other features, we can conclude that the difference between perovskites and pyrochlores can be encapsulated by the two linear models.
For all seven models using the combined database, the normalized RMSE for the pyrochlores is larger than that of the perovskites. On average, when normalized using the standard deviation, the normalized RMSE is 14\% larger for perovskites. This is largely due to significantly larger standard deviations for the pyrochlores (292 meV) than the perovskites (185 meV).

\begin{figure*}[htbp]
 \centering\includegraphics[width=\wholefigwidth]{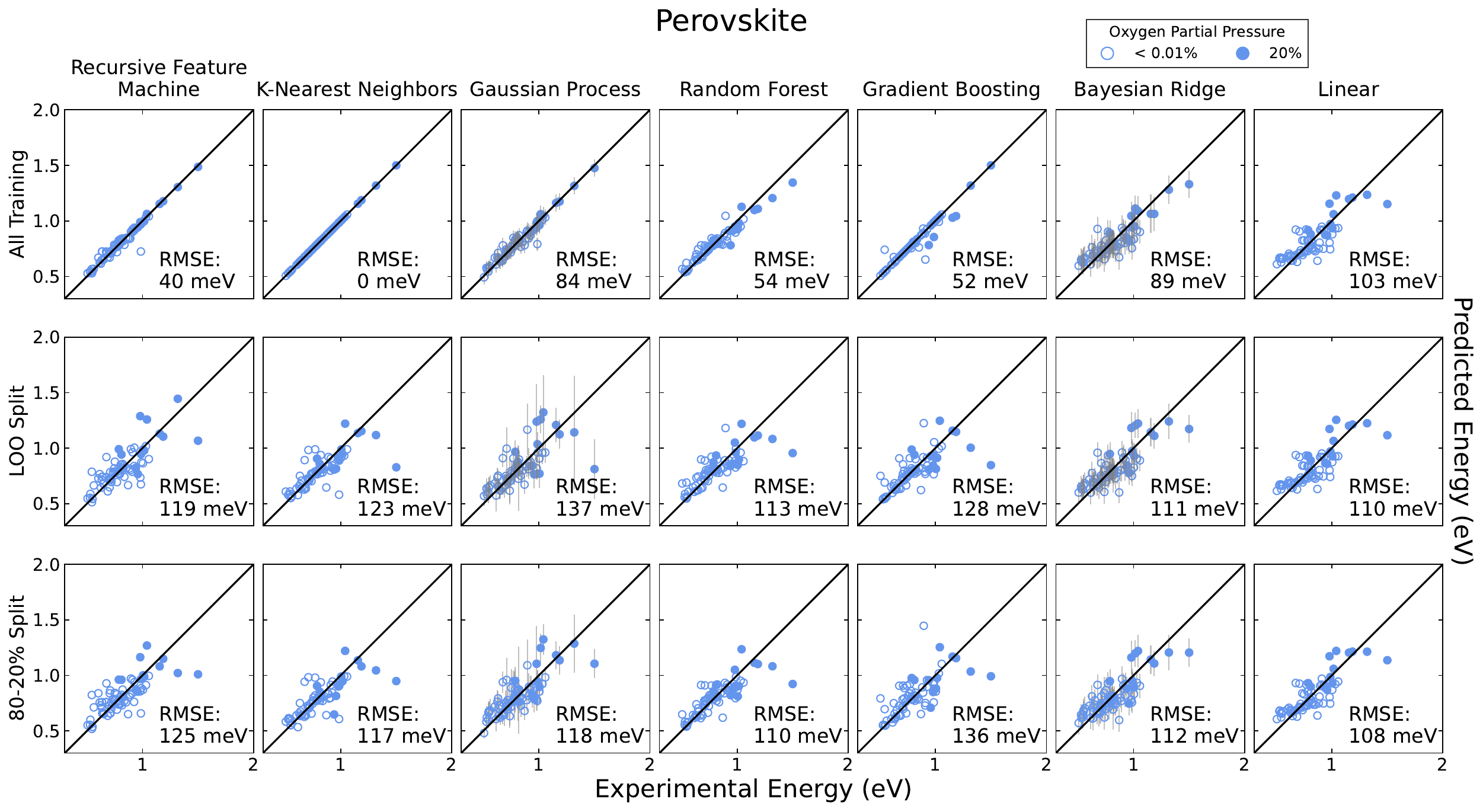}
 \caption{Predicted vs. experimental activation energies for oxygen in perovskites. The marker shape denotes the partial pressure of oxgyen ($p_{\text{O}_2}$) of the experimental measurements. In the middle and bottom rows, we plot only predictions on the test set over all cross-validation splits, and no training data is shown. Our models have RMSEs (199 meV) for activation energies measured at high $p_{\text{O}_2}$  (20\%) that are almost twice as large as those measured at near vacuum (107 meV), likely due to the smaller amount of data at high $p_{\text{O}_2}$ . More complex models, like the Gaussian process or gradient boosting trees, perform the worst on this dataset as their higher flexibilities lead to overfitting.}
 \label{fig:perov_models}
\end{figure*}

After splitting our database by crystal structure, our perovskite models in \Fig{perov_models} show that the most complex models, gradient boosting and Gaussian Process, performed worse than the others, signifying a linear relationship between the features and the activation energy. In fact, the linear model has the best test RMSE for the 80--20\% split, showing that high flexibilities lead to overfitting. The linear model also only uses five features when making predictions, implying that the other ten features could be neglected without harming the accuracy of our predictions. These five features are the Ionic Bonding A, $p_{\text{O}_2}$ , Dev Heat Capacity Mass B, Mean Ground State Volume B, and Mean Covalent Radius A ordered by the absolute value of the coefficients from the linear model. If we split the perovskite database into two parts based on low $p_{\text{O}_2}$  (\textless 0.05\%) or high $p_{\text{O}_2}$  (20\%), we find that oxygen activation energy predictions are worse for high $p_{\text{O}_2}$  measurements, and in fact, four of the seven models (RFM, KNN, RF, and GBR) has higher RMSEs than the standard deviation (189 meV) for high $p_{\text{O}_2}$  activation energies. This is likely due to less data for the high $p_{\text{O}_2}$  subset, as 79\% of the perovskites in our database were measured at low $p_{\text{O}_2}$  and the larger variation in the activation energies. 
This may suggest that at high $p_{\text{O}_2}$ values, the diffusion mechanism may differ; while we take care to only include oxide ion conductors in our dataset, the increase in $p_{\text{O}_2}$ leads to an increased concentration of holes, and thus, the measured diffusivities can be picking up on increased hole conduction. Changes in diffusion mechanisms could be connected to changes in dominant materials features; as we have downselected the important features based on the entire database, we expect our models to have difficulty switching between different mechanisms in their predictions.

\begin{figure*}[htbp]
\centering\includegraphics[width=\wholefigwidth]{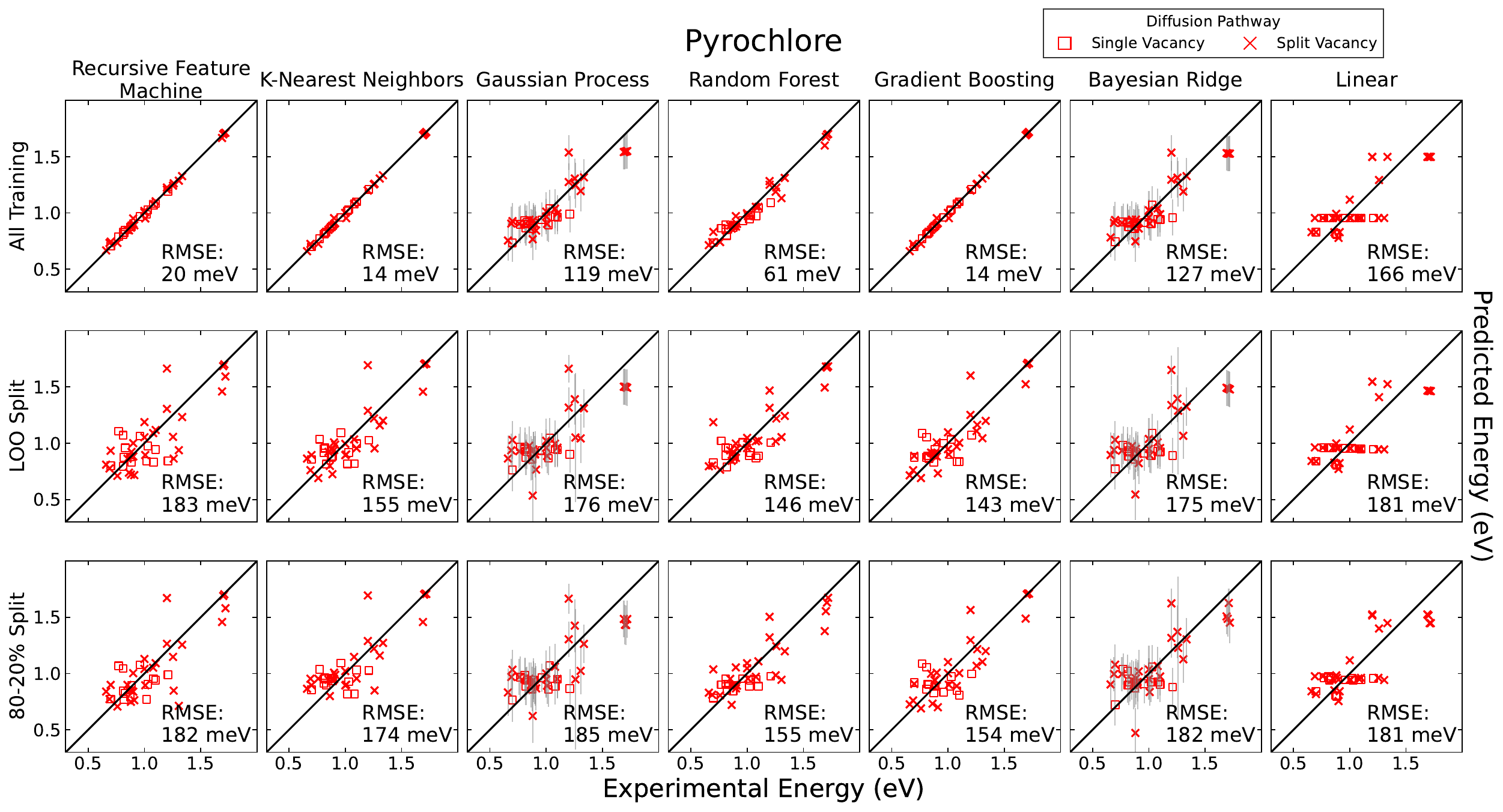}
 \caption{Predicted vs. experimental activation energies for oxygen in all pyrochlores. In the middle and bottom rows, we plot only predictions on the test set over all cross-validation splits, and no training data is shown. The marker shape denotes the most probable dominant diffusion pathway, which is determined by calculating the average ionic radius and comparing it to results from Li and Kowalski \cite{LI2018255},which compared the DFT-calculated defect formation energies of the split and single vacancy. Even though the single vacancy standard deviation (318 meV) is more than double the standard deviation of the split vacancy (142 meV) pyrochlores, the split vacancy RMSEs were almost always larger, except for the linear model. For all models, the testing RMSE of a model trained solely on the split vacancy pyrochlores was larger than the standard deviation.}
 \label{fig:pyro_models}
\end{figure*}

The pyrochlore models fit in \Fig{pyro_models} show that the random forest and gradient boosting trees have RMSEs that are at least 19 meV smaller than the other five models. Decision trees are constructed such that the decision boundaries only depend on a single variable, and so they tend to have smaller RMSEs when the predictions are axis-aligned \cite{axisrf}. The oxygen diffusion in pyrochlores, therefore, should not have a strong dependence on combinations of our chosen features, but instead on the features themselves. 
We note that our representative feature algorithm also reduce the amount of collinearity between our features, which also improves the performance of decision trees.
The two linear models, Bayesian Ridge and linear model, likely lack accuracy due to their lack of flexibility, implying that the relationship between the activation energies and features is non-linear. Similarly, the linear models also have the largest all-training errors. While the kernel-based methods (RFM, GPR, and BR) have large test errors likely due to overfitting, the GPR has similar training RMSE to the two linear models, despite its non-linearity and flexibility. The kernel form we use for the GPR consists of a White and radial basis function (RBF) kernel, where the White kernel considers random noise and the RBF fits a smooth function with a characteristic length scale, which can be roughly described as the distance that one needs to move before the function's value changes significantly. \cite{Rasmussen_Williams_2006} A large length scale would therefore represent a less-flexible model. The length scale for the pyrochlore model is 30 times larger than that of the perovskites, and thus, the GPR model is too inflexible to accurately represent the relationship between the features and the oxygen activation energy.
Our models also find that Ti-containing pyrochlores have larger activation energies (1.5 eV) than the Zr-containing counterparts (0.9 eV). 
Studies have shown that titanium's small ionic radius causes the formation of distorted TiO$_6$ octahedron in the pyrochlore structure, 
both in a high-entropy pyrochlore \cite{Jiang2021}, and in Nd$_2$Ti$_2$O$_7$ through the Jahn-Teller effect \cite{SCHEUNEMANN1975}. 
These distorted octahedrals have been shown through previous experiments to impede oxygen migration 
\cite{ZHONG2018130}, with larger distortions also leading to larger activation energies. 
We further analyze our results based on the theorized diffusion mechanism (single or split vacancy). 
For the split vacancy, the oxygen vacancy lies halfway between two neighboring 48f sites \cite{VANDIJK1985159} instead of localizing to a single 48f site as in the single vacancy case. Previous DFT calculations have suggested the favorability of the split vacancy can be determined using the ionic radii for undoped pyrochlores. Pyrochlores with a more energetically favorable split vacancy have either Zr or Hf on the B site, and an A$^{3+}$ ionic radius that is less than 1.06 \AA \cite{LI2018255}. 
We find that for the split vacancy pyrochlores, the RMSEs are larger than the standard deviation and the RMSEs of the single vacancy pyrochlores, despite the single vacancy pyrochlores having more than double the standard deviation (318 meV) than their split vacancy counterparts (142 meV). This poor behavior is likely due to our feature set being unsuitable for predicting diffusion in split vacancy pyrochlores, as the maximum correlation with the activation energy for any feature was 0.42 (Mean Heat Capacity Molar B). This is in contrast with the single vacancy pyrochlores, which has a much larger maximum correlation of 0.76 (Mean Heat Capacity Mass B).
Finally, we observe large amounts of ``banding'' where the models' predicted values span a much smaller range than the true values. It is the strongest for the linear model where 22 out of 40 pyrochlores have an erroneous prediction between 0.94 and 0.98 eV.

\begin{figure*}[htbp]
\centering\includegraphics[width=\wholefigwidth]{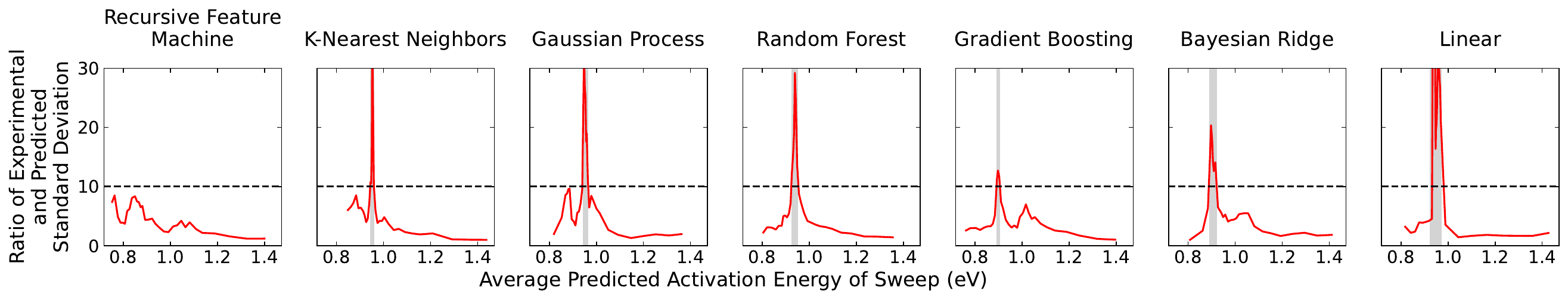}
 \caption{Rolling standard deviation of the experimental activation energies divided by the predicted rolling standard deviation with a width of 8. We identify ``bands'' when the ratio is larger than 10, shown with the grey shading. Banding occurs when a model uses a small number of features that don't vary while the predictions show large variation. The linear model has the strongest banding with a ratio of more than 100 due to its inclusion of only two features (Mean Heat Capacity B and Molar Heat Capacity B). This band consists of all pyrochlores that contain Zr or Sn on their B site. For all models except the RFM, a band is present for predictions between 0.89--0.95 eV.}
 \label{fig:banding}
\end{figure*}

Strong banding occurs in the pyrochlore models, especially for the linear, KNN, and Bayesian Ridge models, as shown in \Fig{banding}, which can be quantified by calculating the ratio of the rolling standard deviation of the predicted activation energies for the testing data from the 80--20\% split with the experimental values. We define a band to occur when the ratio is greater than 10 using a window width of 8. The larger this ratio, the less variability appears, and the more pronounced the banding effect is. The banding occurs when the features being used by the model do not have enough variation, even though a large variation appears in the results. Almost all of the models have a band in the predictions that appears from 0.89--0.95 eV except for the recursive feature machine. This banding is the strongest (ratio of 115) for the linear model, which was limited to only using the two features (Mean Heat Capacity Mass B and Molar Heat Capacity Molar B), both of which only include information about the B-site element. 
This banding is likely caused by the database imbalances as only six elements appear on the B-site, the majority of which contain Zr. 
The band in the linear model consists of all Zr-containing pyrochlores and the single Sn-containing pyrochlore. Thus, the B-site element can be used to roughly split the dataset into small and large activation energies. The more linear models (linear, Bayesian Ridge) are unable to pick up on this relationship, implying that the A-site elemental properties have a nonlinear relationship with the activation energy.

 \begin{figure*}[htbp]
\centering\includegraphics[width=\wholefigwidth]{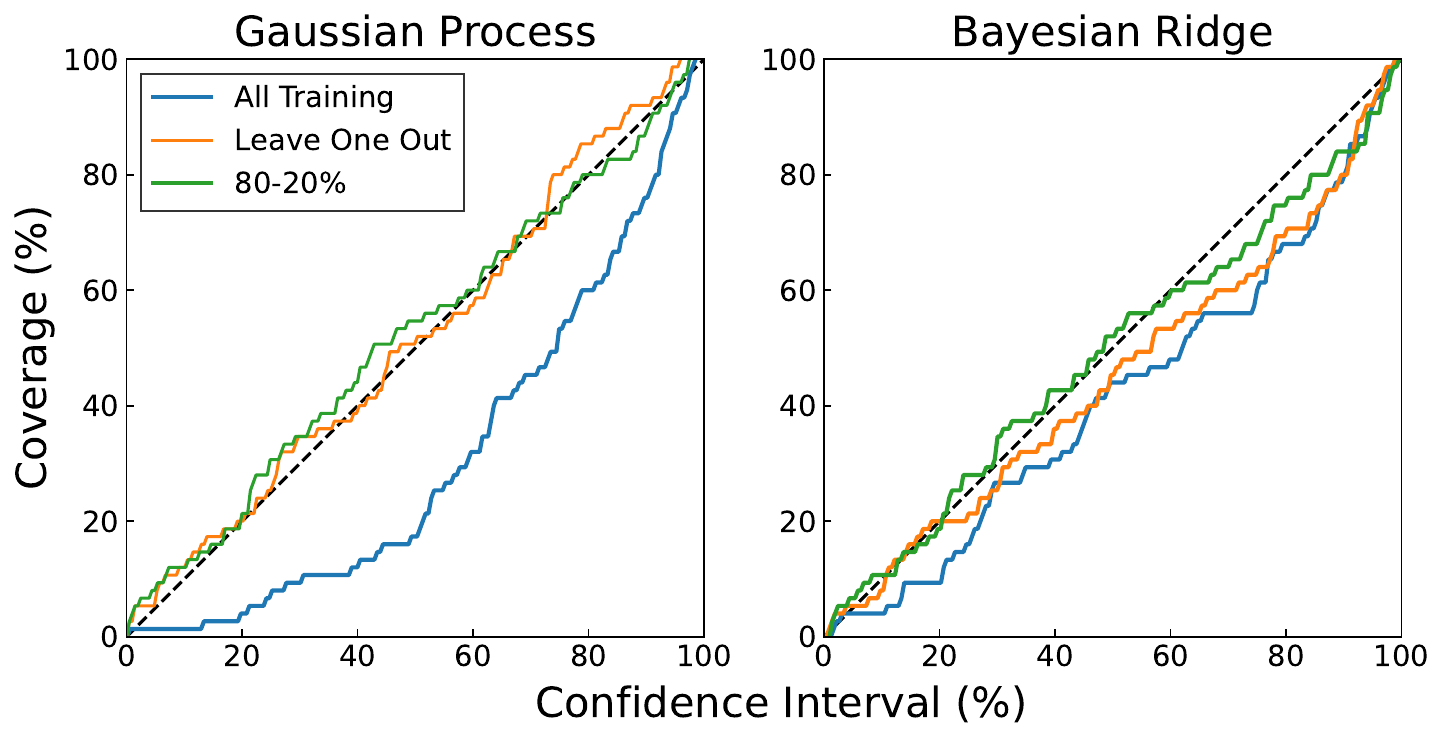}
 \caption{Uncertainty quantification verification on the testing data for the Gaussian process and Bayesian Ridge perovskite models using the coverage percent. The coverage percent denotes the fraction of observed model errors within a given confidence interval. The dashed line corresponds to when the coverage percent matches the confidence interval. Curves that lie below this dashed line denote models that overestimate the error, while curves below the line indicate that the model is underestimating the error. While the all training Gaussian process underestimated the error, we find that all other models accurately predict errors.}
 \label{fig:perov_errors}
\end{figure*}

\begin{figure*}[htbp]
\centering\includegraphics[width=\wholefigwidth]{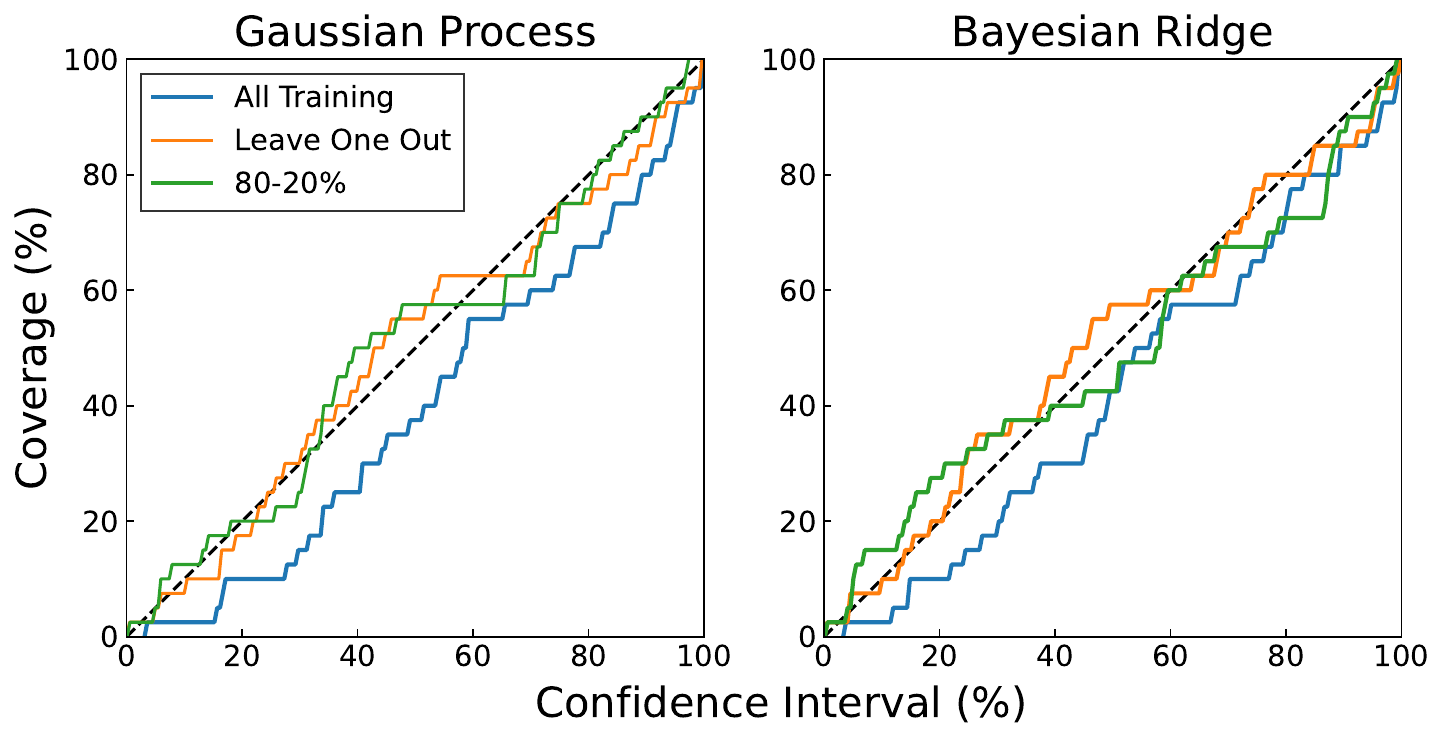}
 \caption{Uncertainty quantification verification on the testing data for the Gaussian process and Bayesian Ridge pyrochlore models using the coverage percent. The coverage percent denotes the fraction of observed model errors within a given confidence interval. The dashed line corresponds to when the coverage percent matches the confidence interval. Curves that lie below this dashed line denote models that overestimate the error, while curves below the line indicate that the model is underestimating the error. While the all training Gaussian process underestimated the error, we find that all other models accurately predict errors.}
 \label{fig:pyro_errors}
\end{figure*}

We perform uncertainty quantification verification on the predicted uncertainty for both the Gaussian process and Bayesian Ridge models as shown in \Fig{perov_errors} and \Fig{pyro_errors} and find that the predicted uncertainty matches the observed errors, particularly on test data. Unfortunately the Gaussian process has large RMSEs compared to the other models, so there is a trade off between native uncertainty predictions and accuracy of the model. For these two models, instead of fitting a single function to the data, they fit a posterior distribution of possible functions, and the standard deviation is an uncertainty prediction. 
We analyze the accuracy of these uncertainty predictions using the coverage percent, which indicates the percent of observed model errors that fall within a given confidence interval. The dotted line corresponds to a perfect match between the coverage percent and the confidence interval. Curves that fall below this line show models that overestimate the uncertainty, while curves above the line show models that are too confident and underestimate the uncertainty. 
In both the perovskite and pyrochlore models, the predicted uncertainty increases as more information is withheld from the training set.
For the perovskites, the average predicted standard deviations increases from 101, 107, and 112 meV for the Bayesian Ridge model and 67, 106, and 188 for the Gaussian Process for the all training, leave one out, and random 80--20\% split respectively.
For the pyrochlores, the average predicted standard deviations is 163, 180, 191 meV for the all training, leave one out, and random 80--20\% split for the Bayesian Ridge, and 173, 183, and 186 meV for the Gaussian Process. For the Gaussian process, for both perovskites and pyrochlores, we find that the models overestimate the all training uncertainty, generally due to larger all training predicted errors despite smaller actual errors. For both perovskites and pyrochlores, the Gaussian process is too flexible and has amongst the largest RMSEs on the 80--20\% testing data. The Bayesian Ridge model performs well on the perovskite database, but not on the pyrochlore database. These small RMSEs only occur when the relationship between the features and activation energy is linear. Thus, for this work, we see tradeoffs between accurate predictions and native uncertainty predictions.


\subsection{Model Explainability}
\label{sec:model_explain}

\begin{figure*}[htbp]
\centering\includegraphics[width=\wholefigwidth]{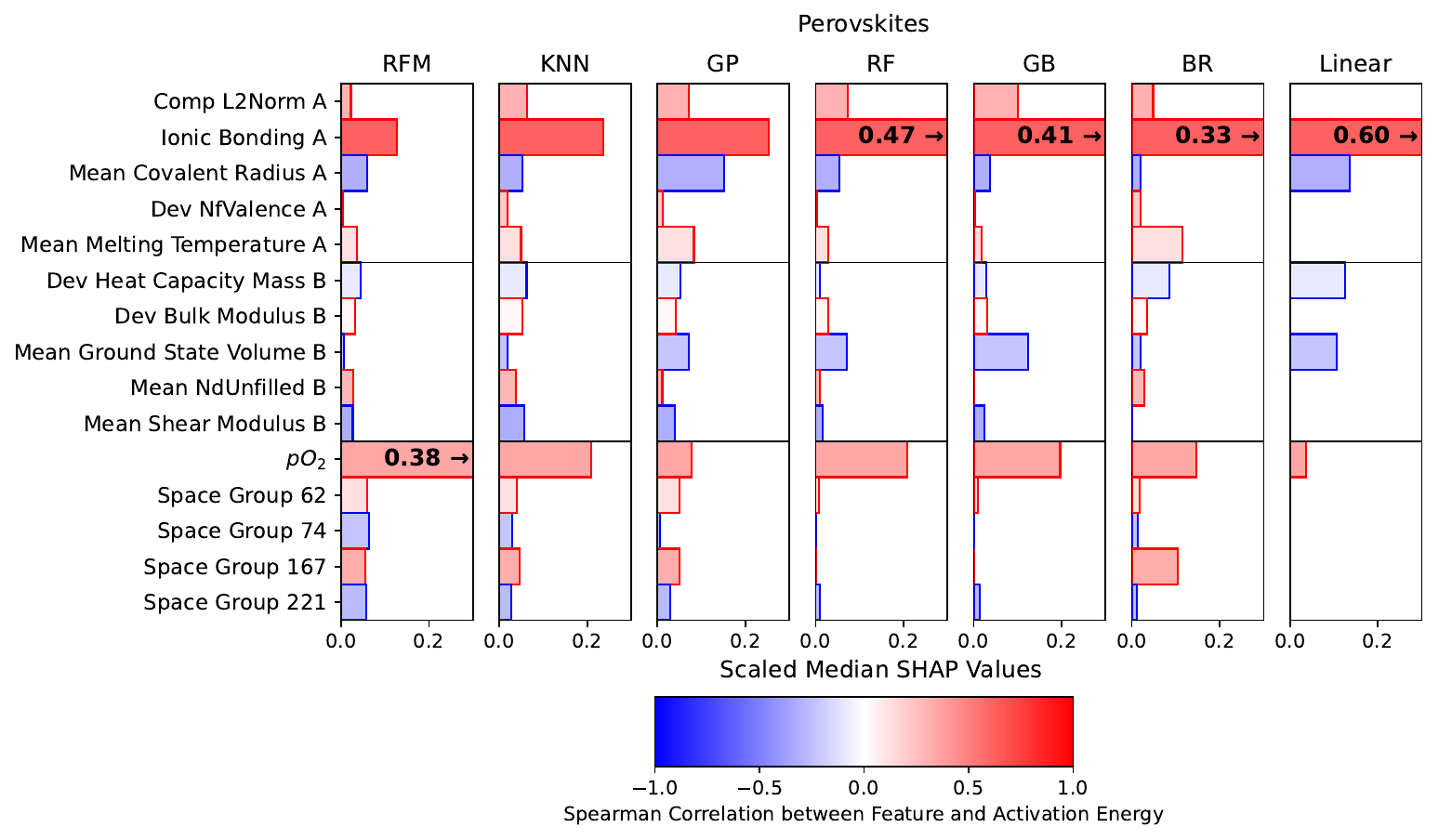}
 \caption{The feature importances for all seven perovskite models. Even though site B is closer to the oxygen site, site B features does not have significantly larger importances than their site A counterparts. All seven models agree that the two most important features are the Ionic Bonding A and $p_{\text{O}_2}$. Ionic Bonding A is the only feature that explicitly combines A and B site elemental properties, while $p_{\text{O}_2}$  measures the experimental conditions.}
  \label{fig:shap_perov}
\end{figure*}

\begin{figure*}[htbp]
\centering\includegraphics[width=\wholefigwidth]{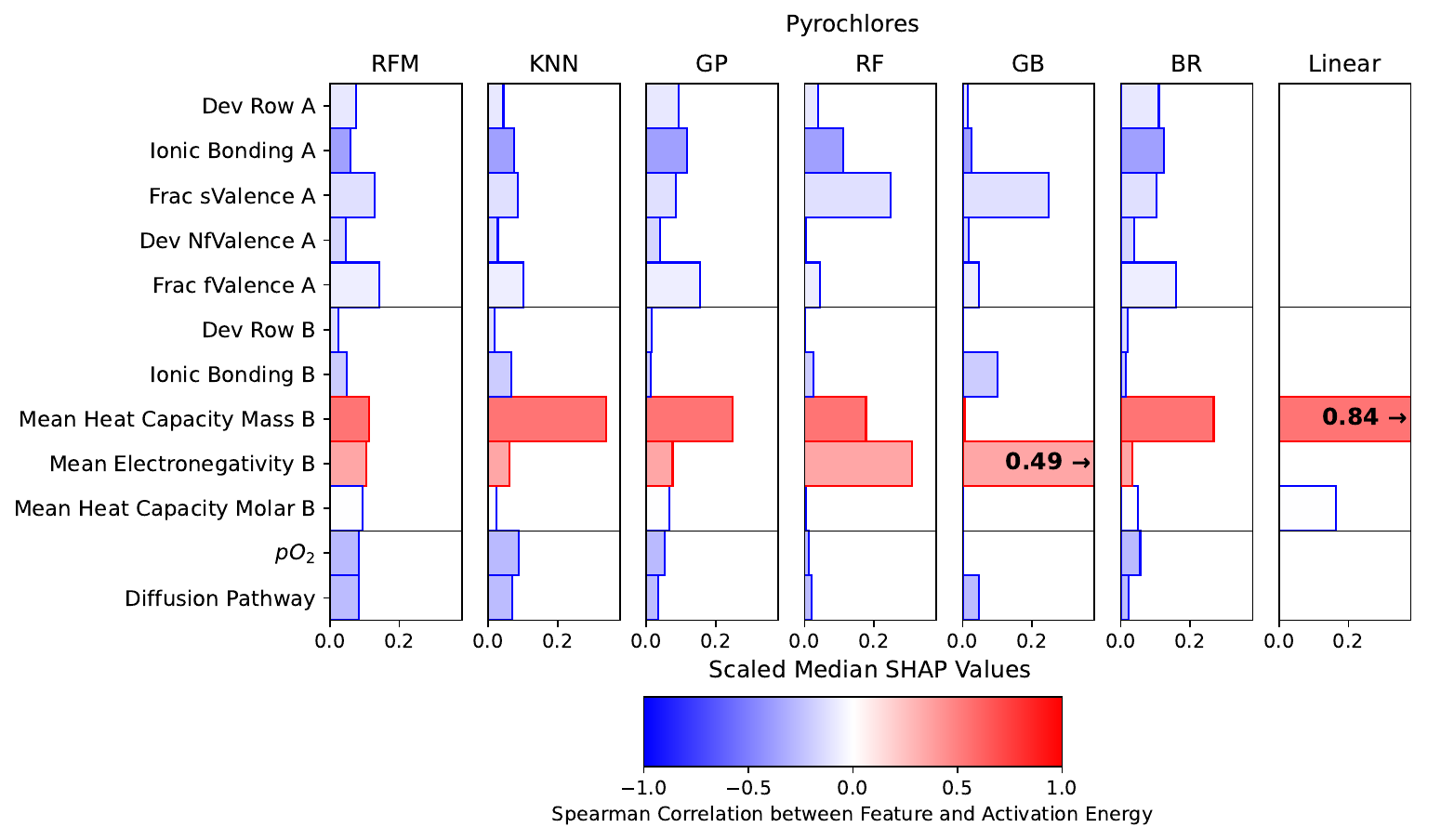}
     \caption{The median SHAP values importances for all seven pyrochlore models. There is less agreement amongst the seven models, but the two (RF and GBR) tree-based models, with the smallest RMSEs, select Mean Electroengativity B and Frac sValence A as the most important features. Both the correlation between the feature and the activation energy and the size of the residual have no bearing on which features are important, pointing towards non-linear relationships between the features and the activation energy. Mean Electronegativity B feature increases the predicted activation energy for Ti-containing pyrochlores, which have an average activation energy of 1.51 eV compared to the 0.94 eV of the other pyrochlores that do not contain Ti.}
     \label{fig:shap_pyro}
\end{figure*}

For all seven models, we calculate the feature importances shown in \Fig{shap_perov} and \Fig{shap_pyro} using the median marginal SHapley Additive exPlanation (SHAP) value \cite{shap}. The SHAP values are calculated by replacing the original model with an interpretable, additive approximation. These values decompose each final prediction into parts determined by a single feature and can take into account correlations between features as it fit a new approximation with every combination of features. A new formulation of the SHAP value \cite{aas2021explaining} has been developed specifically for highly correlated features; however, using that method, features that are not used by the model can have non-zero SHAP values if they are correlated with another feature \cite{sundararajan2020many, janzing2020feature}. Because the grouping algorithm that we use reduces the correlation between features compared to the original feature set, the marginal SHAP value is preferred. The largest correlation is between the Mean Melting Temperature A and Space Group 167 for perovskites (0.81), but both features have low correlations with the activation energy and have low feature importances. For the RFM, there is an inbuilt feature importance method through analyzing the recursively-learned feature matrix $M$ and its eigenvectors to determine which features are being weighted more heavily by the model. For the three datasets, the largest eigenvalue of $M$ is 2 times larger (combined perovskite and pyrochlore), 5 times larger (pyrochlores), or 30 times larger (perovskites) than the second largest eigenvalue. This implies that for predicting oxygen activation energies for perovskites and pyrochlores separately, our features could be projected onto a single vector. While we could analyze both the SHAP value and the eigenvector to find which features are important, they measure two different things. The magnitudes of the features in the largest eigenvector are determined by how the model makes predictions, but the SHAP values only consider the resultant prediction. As the SHAP value could be used for all seven models, we use the SHAP value for direct comparison. For the grouped feature importances from the RFM model, we report the largest SHAP feature importance for any feature within a group. 

For perovksites in \Fig{shap_perov}, the two most important features are Ionic Bonding A and $p_{\text{O}_2}$ . Other than the RFM, none of the B-site features are more than 10\% of the total importance, despite the B-site in perovskites being closer to the oxygen site. Ionic Bonding A has the highest correlation with the activation energy, while the $p_{\text{O}_2}$ has the highest correlation with the residual of a linear model fit between Ionic Bonding A and the activation energy. These two features are the first two to be selected by our greedy algorithm.
In addition to the linear model producing the best fit, we see that the linear model has similar feature importances compared to all other models, except the RFM. This implies that the other models use the features like the linear model does, so the feature-property relationship is likely linear. 
The RFM has similar feature importances compared to the other models, but with less sparseness as it uses all possible features without downselection.
In the largest normalized eigenvector of $M$, only two components have a magnitude larger than 0.5: Ionic Bonding A and $p_{\text{O}_2}$ features. All other components have a magnitude of less than 0.15. Likewise, the two features with the largest median SHAP value are $p_{\text{O}_2}$  and the Ionic Bonding A. 
The Ionic Bonding A is one of the few features that combines both A and B features because one of its terms directly measures the electronegativity difference between the A and B site constituents. It suggests that instead of focusing on specific A- or B-site elements to predict fast or slow oxygen diffusion in perovskites, it would be more useful to look at their differences. This agrees with Ward et al.\ who originally proposed the MAGPIE feature set. After fitting a quadratic function between each MAGPIE descriptor and the DFT-computed formation energy, they found that for materials with at least one nonmetal element, the feature with the smallest RMSE was the Ionic Bonding \cite{Ward2016}.
While the Ionic Bonding A is a feature of the material, $p_{\text{O}_2}$  is an experimental environment feature. For the high $p_{\text{O}_2}$  measurements, the activation energy for perovskites was larger (189 meV) compared to the the low $p_{\text{O}_2}$  values (136 meV). Higher $p_{\text{O}_2}$  values also coincided with a smaller oxygen vacancy concentrations. While the oxygen vacancy concentrations for doped perovskites depends most heavily on the doping level, for undoped materials, the oxygen vacancy depends strongly on the oxygen partial pressure \cite{CHEN1997}.

In \Fig{shap_pyro}, we find that there is much weaker agreement between models for pyrocholores. The RF and GBR models select the Frac sValence A and Mean Electronegativity B as the most import features, while the KNN, GPR, BR, and linear models have Heat Capacity Mass B as the most important feature. The RFM's two most important features, uniquely, are the Frac fValence A and Mean Heat Capacity Mass B. For the RFM, while the first eigenvalue is significantly larger than the others, the corresponding eigenvector does not strongly select specific features. Nine features have a magnitude larger than 0.2, and no feature has a importance larger than 0.4. The SHAP values also show this lack of feature selection, with no specific group selected. Because the RMSEs of the two tree-based models are 20--30 meV smaller than the other models, the Frac sValence A and Mean Electronegativity B feature are the best features for predicting oxygen diffusion in pyrochlores. These two features have a small Spearman correlation (0.12 and 0.36 respectively) with the activation energy itself. Frac sValence A has a piecewise relationship with the activation energy, and can roughly be split into two separate linear functions (above and below a Frac sValence A of 0.45). Such a relationship can be easily represented by a tree-based model as each node splits the data, but other models can only represent this relationship through a more complex nonlinear function. 
The B-site feature (Mean Electronegativity B, Heat Capacity Mass B, or Mean Ground State Volume B) likely help differentiate the Ti-containing pyrochlores, which hasapproximately 0.6 eV higher activation energies than non-Ti containing pyrochlores. Titanium is the element with the smallest electronegativity and largest mean heat capacity of the elements in our database. The KNN, GPR, BR, and linear models also have much smaller site A importances, so the much larger RMSEs of these models likely arise from their lack of sensitivity towards site A features. It is likely that the relationship between the site A features and the activation energy are more complicated with the B-site features, so the less flexible models are unable to capture this relationship. The three models that have some site A importance are also among those with the smallest levels of banding in \Fig{banding}, as larger B site importances emphasize the `categorical' nature of the predictions. While the Bayesian Ridge is the only other model that has a smaller ratio of the experimental rolling standard deviation with the predicted than the RF, GBR, or RFM models, its banding spans a larger range, and its predictions therefore fall into multiple, close in value, bands.


\section{Discussion}
\label{sec:discussion}

In this paper, we create a database and fit machine learning models to predict oxygen diffusion in perovskites and pyrochlores. Multiple machine learning models predict the logarithm of the conductivity in perovskites, and our models have similar training errors. There are no previous machine learning models for predicting conductivities in pyrochlores. 
Zhang and Xu \cite{ZHANG2022} fit a GPR model with a RMSE of 0.1577 for the log of the ionic conductivity at 800$^\circ$C, which corresponded to an activation energy RMSE of 15 meV. When calculating their RMSE, they averaged predictions made by different random 90--10\% cross-validation splits, mixing their training and testing predictions. 
Thus, their model RMSEs were a poor metric for how well their model performs on unseen data. They found that for a leave-one-out cross-validation scheme, the mean absolute percentage error (MAPE) was 3\%, which is identical to the all training MAPEs that we achieve by the RFM, random forest, and gradient boosting models.
In Priya and Aluru's XG-boost model \cite{Priya2021}, they achieved a training RMSE of 47 meV on the activation energies, which was the same RMSE that we obtain by the gradient boosting tree. For the logarithm of the conductivity, they reported a testing RMSE of 0.24 and a validation RMSE of 0.25. As the average temperature of their conductivity data was 922 K, this could be approximated as an activation energy error of 20 meV. Their models had smaller errors compared to ours likely because their dataset was more homogeneous. In their reported datasets, they included conductivities of the same material measured at different temperatures as separate data points, so there was more overlap between the train and test splits. Our RMSEs, therefore, are a more accurate representation of a prediction for an unseen perovskite.

For our six models, we find that the Ionic Bonding A is the single most important feature for predicting oxygen diffusivity in perovskites.
If we compare our feature importances in \Fig{shap_perov} and other machine learning models for perovksite ionic conductivities, all of the machine learning models agree that features representing the bond character and physical structure are the most important feature in calculating the activation energy, but we find that the bond character is more important than the physical structure.
Zhang and Xu (2022) \cite{ZHANG2022} fit a GPR model using six different features. Of these features, five were combinations of the electronegativity and charges and are believed to describe the electronic structure of the BO$_6$ octahedron. Similarly, Priya and Aluru's model \cite{Priya2021}, which used a more comprehensive feature set, found that the minimum electronegativity of site B elements was the most important feature for predicting perovskite conductivities. These electronegativities likely served as a measure of the ionic character of the bonds, specifically between oxygen and other metal atoms.
As the Ionic Bonding A increased (and thus, the ionicity of the bonds with A-atom elements), so did the average activation energy. 
This was surprising as we would expect that the B-site bond strength was more important than the A-site due to its closer proximity to the O-atom. The strong importance of A-site electronegativities was supported by previous work that showed that A-site electronegativities could be used to screen for HER-active perovskites, despite the active site being the B-site atom \cite{Guan2019}.
The higher importance for the A-site could also be due to smaller deviations in Ionic Bonding B due to less variety among B-site constituents in our experimental database. 
The terms for calculating Ionic Bonding A as shown in \Eqn{ionicbonding}, ordered from largest magnitude to the smallest, are the A-O, A-B, and the A1-A2 bond. The largest term, the A-O bond strength, points towards the metal-oxygen bond as being the most important feature, which was supported by DFT calculations performed by Mayeshiba and Morgan \cite{MAYESHIBA2016}, which showed that the oxygen activation energy correlated well with M-O bond strength features, such as the oxygen $p$-band center energy and the vacancy formation energy. Oxygen diffusion in perovskites is highly correlated to the oxygen vacancy formation energies. Creation of an oxygen vacancy involves both the breaking of two B-O bonds and the redistribution of two extra electrons throughout the defected material. DFT calculations have shown that higher electron affinities (and thus, higher electronegativities and smaller iconicity of the B-O bond) stabilize the extra negative charges, reducing both the migration barrier and vacancy formation energies \cite{munoz2014oxygen}. 
While the next two terms are at least an order of magnitude smaller, because of the strong correlation between electronegativity and ionic radii, we can relate the A-B bond strength with the Goldschmidt tolerance factor. The A1-A2 bond is a measure of ionic doping on the A-site, emphasized by the fact that the Ionic Bonding feature group consists of elemental A-site standard deviations instead of weighted averages. While we cannot decompose the effect of increased elemental variation on the A-site sublattice from the stronger effect from the A-O bond strength, most of the doping that occurs is aliovalent in our database, and thus, the A-site doping leads to an increased number of oxygen vacancies and likely increased diffusivities. While it is likely that the increased carrier concentration is the main cause of the increased conductivities, we would be remiss to ignore the effects of short-range order on the cation lattice. A DFT study on the probabilities of A-site cation disorder found that the larger the difference between the cation radii, the more likely it is for A-site ordering to occur in AA'B2O5 perovskites \cite{Albrecht2023}. Due to the strong correlation between the cation radii and the electronegativity, it is possible that the Ionic Bonding A also includes the likelihood of ordering of A-site cations.

The partial pressure of oxygen did not appear as an important feature in previous work on ML models for oxygen diffusion in perovskites despite being the second most important feature for our work. Priya and Aluru \cite{Priya2021} did include it as one of 111 features, but it had negligible importance for their models (the maximum importance it did have was 0.5\% of the total importance in characterizing the primary carrier species). This may be because they have used no method of feature reduction in their work, so strong correlations with other features may have masked the importance of the $p_{\text{O}_2}$. In this work, because there are no cases where the same perovskite is included in our database with two different partial pressures, it's possible that the $p_{\text{O}_2}$  importance is due to its correlation with other material properties. The feature it is the most highly correlated with (0.5 Pearson correlation) is Mean Atomic Weight A, which is in the same feature group as Ionic Bonding A. 
Even though we screen our database to only include materials where the dominant charge carriers are oxygen ions, it is possible that for perovskites whose ionic conductivities were measured at high $p_{\text{O}_2}$ , the increased concentration of holes led to the conductivities to be measured for a mixed O+E regime \cite{NOMURA1997229}.

We propose that perovskites with fast oxygen diffusion should have A-site elements with electronegativities that are closest to that of oxygen (3.44) and be measured at low oxygen partial pressures. Using the Materials Project database, we filtered for experimentally synthesized cubic perovskites with the ABO$_3$ chemical formula, and found a total of 297 different perovskites. Using the seven models that we trained using the entire database, we predicted activation energies for all the perovskites assuming low $p_{\text{O}_2}$. The two linear models predicted large negative activation energies for 54 perovskites. The smallest predicted activation energy was -9.33 eV for LiTaO$_3$. 
For the final prediction, we take the average prediction over the five nonlinear models, and predict that the ten perovskites with the lowest activation energies are SrSeO$_3$ (0.49 eV), CsIO$_3$ (0.50 eV), BaPbO$_3$ (0.53 eV), HgSeO$_3$ (0.53 eV), HfPbO$_3$ (0.53 eV), SrIrO$_3$ (0.54 eV), NdGaO$_3$ (0.55 eV), CdSeO$_3$ (0.55 eV), PrGaO$_3$ (0.55 eV), and CaRhO$_3$ (0.56 eV). 
The estimated uncertainty has a high negative correlation (--0.56) with the prediction, and unfortunately, smaller activation energies also coincide with larger errors. For SrSeO$_3$, the estimated uncertainty is 0.66 eV. Due to the larger variation in the elements that can form a perovskite, the average estimated error was 0.45 eV. This is likely a result of the limited elemental variety in our database; for the Mean Electronegativity B Feature, the range of values is significantly larger (1.12--2.74) compared to our original dataset (1.54--1.85). Our predictions also have a strong negative correlation of --0.88 with the Mean Electronegativity B, our most important feature, and Se, I, and Sr are the two B-site elements with the largest electronegativities.

For pyrochlores, the two best-performing models, RF and GBR, find a strong dependence on Frac sValence A and Mean Electronegativity B, followed distantly by Ionic Bonding A and Ionic Bonding B.
For the most important feature, Frac sValence A, we find that larger values of Frac sValence A, which corresponds to Ca, Y, or La on the A site, lead to 0.2 eV larger activation energy. These increased migration barriers agree in part with DFT calculations performed by Li and Kowalski \cite{LI2018255}, which found that the oxygen migration barriers between neighboring 48f and 48f sites (single vacancy mechanism) were 0.1 eV larger for (Y,La)$_2$Zr$_2$O$_7$. Unlike their results, however, we find that the effects of these specific A-site elements are larger for Ti-containing pyrochlores.
The second important feature, the Mean Electronegativity B, acts as a surrogate for the ionic radii. Ti, which is the smallest of all possible B site cations, also has a larger electronegativity (1.54). It has previously been shown that in high-entropy pyrochlores, the inclusion of Ti leads to distorted TiO$_6$ octahedrals. This is because its small ionic radii can stabilize the weberite structure compared to the pyrochlore one, and the second order Jann-Teller effect is the strongest for Ti compared to Zr and Hf \cite{jiang2020probing}. These distorted octahedrals impede diffusion. Additionally, in Ti-containing pyrochlores, the Ti cations are 6-fold coordinated, which is optimal, while Zr-containing pyrochlores prefer a 7-fold coordination. In Ti-containing pyrochlores, this locks the oxygen atoms in place as any movement of the oxygen atoms would disrupt the optimal coordination. The frustration of the Zr cations can be temporarily alleviated by oxygen migration, and thus, the oxygen atoms are more free to move throughout the Zr-containing pyrochlore \cite{pilania2019distortion}.
The next most important features, Ionic Bonding A and B, 
can be thought of a measure of how different the electronegativities are between B-site elements, and therefore are a measure of cation disorder on the B lattice. This agrees with classical potential calculations that have shown that increased cation disorder can decrease the formation energy of anion Frenkel defects \cite{minervini}, and experimental studies that found that cation disorder lead to increased conductivities \cite{kreller2019}, and was the largest at the boundary before the ordered pyrochlore structure turns into a more disordered defected fluorite structure \cite{YAMAMURA2003359}. Molecular dynamics simulations with a Buckingham potential done by Perriot and Uberuaga had shown that, in Gd$_2$ZrO$_7$ and Gd$_2$Ti$_2$O$_7$, with increased cation disorder, both the barriers and trap depth decreased, due to weaker bonding of the oxygen \cite{perriot2015}. Unlike the perovskites, we find little impact of $p_{\text{O}_2}$ , with it only having nonzero feature importances for the RFM and KNN feature, which represents the lack of variety in experimental $p_{\text{O}_2}$  for pyrochlore oxygen diffusion measurements. Only two pyrochlores were measured at low $p_{\text{O}_2}$ , while all others were measured at standard atmosphere conditions ($p_{\text{O}_2}$ = 20\%). 

We propose that new pyrochlores with fast oxygen ion conductivities should not contain Ti on the B site, have A-site atoms with a fraction of s Valence electrons between 0.2--0.4, and ideally have cation disorder. Specifically, these new pyrochlores should have configurations that lie at the boundary between the ordered pyrochlore and the disordered fluorite structure. Using the Materials Project database and screening for all experimentally-synthesized A$_2$B$_2$O$_7$ materials in the 227 space group, we find a total of 115 possible pyrochlores. We assume that the $p_{\text{O}_2}$  is 20\% or standard atmosphere conditions as most pyrochlores in our database were measured at those conditions. Using the RF and GBR to predict the activation energies (these two models are chosen as they have significantly smaller RMSEs compared to the others), we find that no material has a smaller activation energy than Sm$_2$Zr$_2$O$_7$ (0.69 eV), which has already been measured experimentally and is included in our database. The next-smallest predicted activation energy is for Eu$_2$Hf$_2$O$_7$, with a predicted activation energy of 0.82 eV, followed by  Eu$_2$Zr$_2$O$_7$. All 8 Zr-containing pyrochlores have activation energies within the smallest 10 pyrochlores.
Using the GPR model, we estimate the uncertainty for that prediction to be 0.17 eV. Due to less elemental variety than the perovskites, the average estimated uncertainty is half the size (0.28 eV).

In this work, we construct a database of experimental activation energies in perovskites and pyrochlores, and using our feature grouping analysis, we interpret feature importances from ML models on this task. While previous ML models had been fit to conductivities in perovskites, we perform a detailed feature analysis and explore the relative importances of seven different ML models.
Unlike Baldarassi et al. \cite{baldassarri2023oxygen}, we find that the oxide properties were unused in our model, as no oxide property is chosen as the representative feature of its group. Even though Mean Oxide Bulk Modulus A and Mean Oxide Bandgap A were the two features with the largest correlations with the activation energy for the perovskites in their feature group, the chosen feature (Mean Melting Temperature A for perovskites) have a larger correlation with the residual of a linear model fit to the previously-chosen features.
These machine learning models not only enable predictions of transport properties in perovskites and pyrochlores, but also provide physical insights into which features are important. We find that differences in electronegativity, measured by the Ionic Bonding, is within the top three most important features for both perovskites and pyrochlores.The material properties used in this model are easy to measure and calculate, and therefore can enable rapid screening of new oxides for solid oxide fuel cells or other applications. We show that our grouping analysis of feature importances is applicable even in cases where the weighted averages of elemental properties is not an accurate representation of the material of interest's true properties.

\begin{acknowledgments}
This work was supported by by the National Science Foundation Graduate Research Fellowship under Grant No. DGE-1746047 (G.L.). The authors thank Prof. Ansu Chatterjee for the helpful comments and discussions. The data is available as supplemental material \footnote{See Supplemental Material at [URL will be inserted by publisher] for CSV files of perovskite and pyrochlore data.}.
\end{acknowledgments}


%
\end{document}